\documentclass[11pt]{article}
\usepackage{amssymb}
\usepackage{graphics}
\usepackage{epsfig}
\usepackage{a4wide}
\usepackage{slashed}
\usepackage{multicol,bbm}
\usepackage{multirow}

\begin{document}

\title{Possible effects of the large extra dimensions on $ZZW$ production at the LHC }
\author{ Chen Chong, Guo Lei, Ma Wen-Gan, Zhang Ren-You, Li Xiao-Zhou, and Zhang Yu \\
{\small  Department of Modern Physics, University of Science and Technology of China (USTC),}  \\
{\small  Hefei, Anhui 230026, People's Republic of China}}

\date{}
\maketitle \vskip 8mm
\begin{abstract}
We investigate the possible large extra dimensions (LED) effects induced by the Kaluza-Klein gravitons up to the QCD next-to-leading order (NLO) on $ZZW$ production at the large hadron collider (LHC). The integrated cross sections and some kinematic distributions are presented in both the standard model (SM) and the LED model. The numerical results demonstrate that the NLO QCD corrections are sizeable and remarkably reduce the leading order (LO) LED effect depending strongly on the phase space. The NLO LED relative discrepancies of the total cross section could become sizable for the $ZZW$ production, if we apply proper event selection criteria. We find that the LO result overestimates the LED effect and is insufficient to provide a believable theoretical prediction.
\end{abstract}


\vfill \eject \baselineskip=0.32in

\renewcommand{\theequation}{\arabic{section}.\arabic{equation}}
\renewcommand{\thesection}{\Roman{section}.}
\newcommand{\nb}{\nonumber}

\newcommand{\Dir}{\kern -6.4pt\Big{/}}
\newcommand{\Dirin}{\kern -10.4pt\Big{/}\kern 4.4pt}
\newcommand{\DDir}{\kern -7.6pt\Big{/}}
\newcommand{\DGir}{\kern -6.0pt\Big{/}}

\makeatletter      
\@addtoreset{equation}{section}
\makeatother       

\par
\section{Introduction}
\par
To solve the long-standing hierarchy problem, many exciting extensions of the standard model (SM) have been developed,
such as supersymmetric models, little Higgs models, extra dimension models, and etc. Among these extensions, the large
extra dimensions (LED) model proposed by Arkani-Hamed, Dimopoulos and Dvali \cite{LEDM-1, LEDM-2, LEDM-3} is an attractive one because it
predicts possible quantum gravity effects at TeV scale. In the LED model, we have $(4+d)$-dimensional spacetime with
$d$ being the number of extra spatial dimensions compactified on a $d$-dimensional torus with radius $R$.  The SM
particles are confined on a $4$-dimensional brane world while the graviton can propagate in the $(4+d)$-dimensional
bulk. The Planck scale $M_P$ in the $4$-dimensional spacetime is related to the fundamental scale $M_S$ of the LED model
as $M_P^2 = R^d M_S^{d+2}$. If $R$ is large enough, the gravity interaction governed by $M_S$ and the gauge interaction
can be unified at the TeV scale, therefore, the gauge hierarchy problem is solved.

\par
In the low-energy effective theory of the LED model, the massless graviton propagating in the $(4+d)$-dimensional spacetime
is equal to a tower of massive Kaluza-Klein (KK) states only propagating in the ordinary $4$-dimensional spacetime \cite{LED-KK}.
After performing the KK reduction, we obtain the $4$-dimensional interactions of the SM particles with the KK gravitons.
These effective couplings are heavily suppressed by $1/\overline{M}_P$, where $\overline{M}_P$ is the reduced Planck scale
defined as $\overline{M}_P = M_P/\sqrt{8 \pi}$. However, for both the virtual graviton exchange and the real graviton
production, the summation over the tower of KK states cancels the dependence on $\overline{M}_P$ and leads to a suppression
of the order of $M_S$. Therefore, the KK-graviton effects may be considerable at high energy colliders.

\par
The triple gauge boson (TGB) productions are of particular interest because they are sensitive to the quartic gauge couplings
(QGCs) and thus related to the electroweak symmetry breaking (EWSB) mechanism \cite{QGC-1, QGC-2}. Any deviation from the SM predictions would
hint at the existence of new physics. Therefore, the precision studies on the TGB productions at high energy colliders within
all the possible new physics models are necessary in discriminating physics beyond the SM. All the TGB productions at hadron
colliders have been studied in the SM up to the QCD next-to-leading order (NLO). It is found that the NLO QCD corrections are
sizable, strongly depend on the phase space and significantly exceed the expectations from a scale variation of the leading
order (LO) result. Therefore, the NLO QCD corrections should be taken into account for the TGB production phenomenological study.

\par
The CERN Large Hadron Collider (LHC) is expected to explore the mechanism of EWSB and new physics evidences at the TeV scale,
and provide more precision measurements of the QGCs than the existing data from LEP II and Tevatron \cite{QGC-collider-1, QGC-collider-2}. Compared with
the thoroughly studied TGB productions at hadron colliders in the SM, only the
$pp \to \gamma\gamma\gamma,~ \gamma\gamma Z,~ \gamma ZZ,~ ZZZ,~ W^+W^-\gamma,~ W^+W^-Z$ processes at the LHC were
studied at the LO in the framework of the LED model \cite{TGB-1, TGB-2}. As the $ZZW$ production is directly related to the $ZZW^+W^-$
QGC, the precision theoretical predictions on the $ZZW$ production are needed for the measurement of the $ZZW^+W^-$ coupling
and the search for the new physics signature in experiment.
Recently, the generalized effective $W$ approximation is applied to study the $WW$ scattering based on the factorization \cite{WW-scattering}.
In this paper, we investigate the possible LED effects on the
$ZZW$ production at the CERN LHC up to the QCD NLO. The rest of the paper is organized as follows. In Sec. II, the calculation
strategy is presented. In Sec. III, the numerical results and discussions are provided. Finally, a short summary is given
in Sec. IV.

\par
\section{Calculation strategy}
\par
\subsection{Related LED theory }
\par
In this work we adopt the de Donder gauge and Feynman gauge for the KK-graviton part and SM part, respectively.
The related Feynman rules used in our calculations are listed in Table \ref{tab0},
where $G_{\rm KK}^{\mu \nu}$, $\psi$, $A^{a \mu}$, $W^{\pm \mu}$ and $Z^{\mu}$ represent the fields of the graviton,
quark, gluon, $W$-boson and $Z$-boson, respectively. The momenta of KK graviton and gauge bosons are set to be into
vertex, while the quark momentum is defined in the direction of the quark flow \cite{LED-KK}. $\alpha_3$ and $\xi$ are
the $SU(3)$ and charged $SU(2)$ gauge fixing parameters which are taken as $\alpha_3 = \xi = 1$ in Feynman gauge.
$\kappa = \sqrt{2}/\overline{M}_P$ is the gravity coupling constant. $D(s)$ in the KK-graviton propagator can be
expressed as \cite{LED-KK}
\begin{eqnarray}
\label{Res}
D(s) = \frac{16 \pi}{\kappa^2} \frac{s^{d/2-1}}{M_S^{d+2}}
\biggl[
\pi + 2i I(\Lambda/\sqrt{s})
\biggr],
\end{eqnarray}
where
\begin{eqnarray}
I(\Lambda/\sqrt{s}) = P \int_0^{\Lambda/\sqrt{s}} dy \frac{y^{d-1}}{1-y^2}.
\end{eqnarray}
The integral $I(\Lambda/\sqrt{s})$ contains an ultraviolet cutoff $\Lambda$ on the KK modes \cite{LED-KK, LED-2} and in our
calculations we set $\Lambda$ to be the fundamental scale $M_S$ routinely. The tensor coefficients
$B^{\mu \nu \alpha \beta}$, $C^{\rho \sigma \mu \mu \alpha \beta}$ and $E^{\mu \nu \rho \sigma}(k_{1},k_{2})$ are
given by \cite{tensor-coe}
\begin{eqnarray}
B^{\mu \nu \alpha \beta} &=& \frac{1}{2}
      (\eta^{\mu \nu}\eta^{\alpha \beta}
      -\eta^{\mu \alpha}\eta^{\nu \beta}
      -\eta^{\mu \beta}\eta^{\nu \alpha}),
       \nonumber \\
C^{\rho \sigma \mu \nu \alpha \beta} &=& \frac{1}{2}
      [
       \eta^{\rho \sigma}\eta^{\mu \nu}\eta^{\alpha \beta}
     -(\eta^{\rho \mu}\eta^{\sigma \nu}\eta^{\alpha \beta}
      +\eta^{\rho \nu}\eta^{\sigma \mu}\eta^{\alpha \beta}
      +\eta^{\rho \alpha}\eta^{\sigma \beta}\eta^{\mu \nu}
      +\eta^{\rho \beta}\eta^{\sigma \alpha}\eta^{\mu \nu})
      ],
      \nonumber \\
E^{\mu \nu \rho \sigma}(k_{1},k_{2}) &=&
      \eta^{\mu \nu}(k_1^{\rho} k_1^{\sigma} + k_2^{\rho} k_2^{\sigma} + k_1^{\rho} k_2^{\sigma}) -
      [
      \eta^{\nu \sigma} k_1^{\mu} k_1^{\rho} + \eta^{\nu \rho} k_2^{\mu} k_2^{\sigma} +
      (\mu \leftrightarrow \nu)
      ].
\end{eqnarray}
We code the related Feynman rules of the LED model in the FeynArts-3.5 package \cite{feynarts} to generate
the Feynman diagrams and the corresponding amplitudes. Our developed FormCalc-5.4 package \cite{formcalc} is used subsequently
to simplify the amplitudes. In the following, we present the calculation details of different contribution parts to
the $ZZW$ production at the LO and QCD NLO, and the verification of the consistency of our results with previous
publications \cite{ZZW-check1, ZZW-check2}.
\begin{table}[!hbp]
\small
\renewcommand{\arraystretch}{1.9}
\begin{center}
\begin{tabular}{|l|l|}
  \hline
  ~~~~~~~~~~~~~~ Vertex & ~~~~~~~~~~~~~~~~~~~~~~~~~~~~~~ Feynman rule \\
  \hline
  $G_{\rm KK}^{\mu \nu}(k_3)A^{a\rho}(k_1)A^{b\sigma}(k_2)$ &
  $
  -i \kappa
  \Bigl[
  (C^{\mu \nu \rho \sigma \tau \beta} - C^{\mu \nu \rho \beta \sigma \tau}) k_{1\tau} k_{2\beta} +
  \frac{1}{\alpha_3} E^{\mu \nu \rho \sigma}(k_1, k_2)
  \Bigr] \delta^{a b}
  $ \\
  $G_{\rm KK}^{\mu \nu}(k_3)W^{+ \rho}(k_1)W^{- \sigma}(k_2)$ &
  $
  -i \kappa
  \Bigl[
  (C^{\mu \nu \rho \sigma \tau \beta} - C^{\mu \nu \rho \beta \sigma \tau}) k_{1\tau} k_{2\beta} +
  \frac{1}{\xi} E^{\mu \nu \rho \sigma}(k_1,k_2) + B^{\mu \nu \rho \sigma} m_W^2
  \Bigr]
  $ \\
  $G_{\rm KK}^{\mu \nu}(k_3)Z^{\rho}(k_1)Z^{\sigma}(k_2)$ &
  $
  -i \kappa
  \Bigl[
  (C^{\mu \nu \rho \sigma \tau \beta} - C^{\mu \nu \rho \beta \sigma \tau}) k_{1\tau} k_{2\beta} +
  \frac{1}{\xi} E^{\mu \nu \rho \sigma}(k_1,k_2) + B^{\mu \nu \rho \sigma} m_Z^2
  \Bigr]
  $ \\
  $G_{\rm KK}^{\mu \nu}(k_3)\bar{\psi}(k_1)\psi(k_2)$ &
  $
  -i \frac{\kappa}{8}
  \Bigl[
  \gamma^{\mu} (k_1 + k_2)^{\nu} +
  \gamma^{\nu} (k_1 + k_2)^{\mu} -
  2 \eta^{\mu \nu} (\rlap/{k}_1 + \rlap/{k}_2 - 2 m_{\psi})
  \Bigr]
  $ \\
  $G_{\rm KK}^{\mu \nu}(k_4)\bar{\psi}_{u_i}(k_1)\psi_{d_j}(k_2)W^{+ \rho}(k_3)$ &
  $
  -i e \frac{\kappa}{4 \sqrt{2} \sin\theta_W}
  \left(
  \gamma^{\mu} \eta^{\nu \rho} +
  \gamma^{\nu} \eta^{\mu \rho} - 2
  \gamma^{\rho}\eta^{\mu \nu}
  \right)
  V_{CKM}^{ij}
  $ \\
  $G_{\rm KK}^{\mu \nu}(k_4)\bar{\psi}(k_1)\psi(k_2)A^{a\rho}(k_3)$ &
  $
  i g_{s} \frac{\kappa}{4}
  \left(
  \gamma^{\mu} \eta^{\nu \rho} +
  \gamma^{\nu} \eta^{\mu \rho} -
  2 \gamma^{\rho}\eta^{\mu \nu}
  \right)
  T^{a}
  $ \\
  \hline
  \hline
  \multicolumn{2}{|l|}{Spin-2 KK-graviton propagator after summation over KK states:} \\
  \multicolumn{2}{|r|}{$
  \tilde{G}_{\rm KK}^{\mu \nu \alpha \beta}
  =
  \frac{1}{2} D(s)
  \Bigl[
  \eta^{\mu \alpha} \eta^{\nu \beta} +
  \eta^{\mu \beta} \eta^{\nu \alpha} -
  \frac{2}{d + 2}\eta^{\mu \nu} \eta^{\alpha \beta}
  \Bigr]~~~$} \\
  \hline
\end{tabular}
\end{center}
\begin{center}
\begin{minipage}{15cm}
\caption{\label{tab0} Related LED Feynman rules used in this work.}
\end{minipage}
\end{center}
\end{table}

\subsection{LO cross section}
\par
\begin{figure}[htbp]
 \centering
 \includegraphics[width=0.65\textwidth]{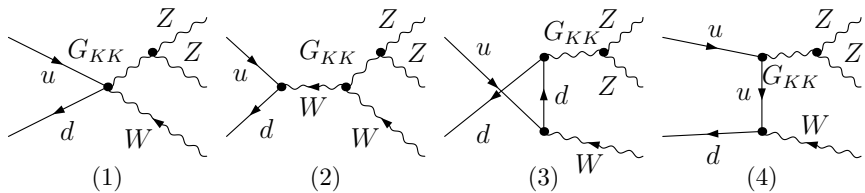}
 \centering
 \caption{The LO Feynman diagrams with KK-graviton exchange for the partonic process $u\bar{d} \to ZZW^+$.}
 \label{Feyndiag}
\end{figure}

We neglect the masses of $u$-, $d$-, $c$-, $s$-quarks and the quark mixing between the third generation and the first two generations. Due to the smallness of the $b$-quark parton density in the proton, only the $qq^{\prime} \to ZZW^+$ $(qq^{\prime}=u\bar{d}, u\bar{s}, c\bar{d}, c\bar{s})$ and $qq^{\prime} \to ZZW^-$ $(qq^{\prime}=\bar{u}d, \bar{u}s, \bar{c}d, \bar{c}s)$ partonic processes are involved in the $ZZW^+$ and $ZZW^-$ productions at the LHC, respectively. At parton level, the cross section for the $ZZW^-$ production is the same as that for the $ZZW^+$ production due to the $CP$ conservation. In this section we present only the analytical results for the $pp \to ZZW^+ + X$ process. The LO Feynman amplitude for the partonic process $qq^{\prime} \to ZZW^+$ can be expressed as
\begin{eqnarray}
{\cal M}^{LO}_{qq^{\prime}} =
{\cal M}_{qq^{\prime}}^{SM} + {\cal M}_{qq^{\prime}}^{LED},
\end{eqnarray}
where ${\cal M}_{qq^{\prime}}^{SM}$ and ${\cal M}_{qq^{\prime}}^{LED}$ are the amplitudes contributed by the SM-like
diagrams and the KK-graviton exchange diagrams, respectively. The LO Feynman diagrams involving KK-graviton exchange
are shown in Fig.\ref{Feyndiag}. The LO cross section for $qq^{\prime} \to ZZW^+$ has the form as
\begin{eqnarray}
\hat{\sigma}^0_{qq^{\prime}} =
 \frac{1}{4 |\vec{p}| \sqrt{\hat{s}}} \int {\sum}^{\prime}
 \left|
 {\cal M}^{LO}_{qq^{\prime}}
 \right|^2
 d \Omega_3,
\end{eqnarray}
where $\vec{p}$ is the three-momentum of one initial parton in the center-of-mass system (c.m.s.), $\sqrt{\hat{s}}$ is
the c.m.s. colliding energy, $d\Omega_3$ is the three-body phase space element, the summation is taken over the spins
and colors of the initial and final states, and the prime on the summation indicates averaging over the intrinsic
degrees of freedom of initial partons. By convoluting $\hat{\sigma}^0_{qq^{\prime}}$ with the parton distribution
functions (PDFs) of the colliding protons, we obtain the LO total cross section for the parent process
$pp \to ZZW^+ + X$ as
\begin{eqnarray}
\sigma_{LO} =
\sum_{qq^{\prime}=u\bar{d}, u\bar{s},}^{c\bar{d}, c\bar{s}}
\int_0^1 dx_1 dx_2
\biggl[
G_{q/P_1}(x_1, \mu_f) G_{q^{\prime}/P_2}(x_2, \mu_f)
\hat{\sigma}^0_{qq^{\prime}}(\sqrt{\hat{s}} = x_1 x_2 \sqrt{s}) + (1 \leftrightarrow 2)
\biggr],
\end{eqnarray}
where $G_{q/P}$ represents the PDF of parton $q$ in proton $P$, $x_i~(i=1,2)$ describes the momentum fraction of a
parton in proton, $\sqrt{s}$ is the colliding energy in the rest frame of proton-proton system, and $\mu_f$ is the
factorization scale.

\subsection{Virtual corrections}
\par
There are 145 QCD one-loop Feynman diagrams for the $qq^{\prime} \to ZZW^+$ partonic process, including 22
boxes and 6 pentagons. These loop diagrams contain both UV and IR singularities. All the UV and part of the IR divergences can be
removed after performing the renormalization procedure by introducing the quark wavefunction renormalization constants
$\delta Z_{q,L}$ and $\delta Z_{q,R}$ which are fixed in the modified minimal subtraction ($\overline{MS}$)
renormalization scheme as
\begin{eqnarray}
\delta Z_{q,L}=\delta Z_{q,L}=
-\frac{\alpha_{s}(\mu_r)}{4 \pi} C_{F} \left( \Delta_{UV} - \Delta_{IR} \right),
\end{eqnarray}
where $C_{F}=\frac{4}{3}$, $\mu_r$ is the renormalization scale, and
$\Delta_{UV}=\frac{1}{\epsilon_{UV}}\Gamma(1+\epsilon_{UV})(4\pi)^{\epsilon_{UV}}$ and
$\Delta_{IR}=\frac{1}{\epsilon_{IR}}\Gamma(1+\epsilon_{IR})(4\pi)^{\epsilon_{IR}}$ refer to the UV and IR divergences
regulated in the dimensional regularization scheme, respectively. The residual IR divergences can be canceled by
adding the contributions of the real emission processes and the corresponding PDF counterterms.

\par
In deduction of the Feynman amplitudes, the 3- and 4-point tensor integrals are recursively reduced to scalar
integrals using Passarino-Veltman (PV) method \cite{PV-method}, while the 5-point integrals are reduced to 4-point
integrals by using the method proposed by Denner and Dittmaier \cite{DD-method}. We should address that the rank $n>3$
tensor 4-point integrals may induce a serious unstable problem in the numerical calculation. One way to solve
this problem is to adopt quadruple precision arithmetic in the numerical calculation of loop integrals, but the
cost is obvious to consume much more computer CPU time. In order to improve the efficiency of the calculation,
we adopt the segmentation method analogous to that in Ref.\cite{unstable-problem-1, unstable-problem-2} to treat the unstable problem.  We developed
the codes for the calculation of the scaler and tensor integrals based on the LoopTools-2.7 package, which can
switch to the quadruple precision codes automatically in the region of
\begin{eqnarray}
\frac{{\rm det} G_3}{(2 k_{max}^2)^3} < 10^{-5},
\end{eqnarray}
where ${\rm det} G_3$ is the Gram determinant and $k_{max}^2$ the maximum of the external four-momentum squared
for a given  4-point integral. The calculation speed by using our modified LoopTools is about ten times faster
than that using pure quadruple precision arithmetic in the whole phase space.

\subsection{Real emission corrections}
\par
We employ the dipole subtraction (DS) scheme proposed by Catani and Seymour to deal with the IR singularities in
the real emission corrections. \cite{dipole-subtraction}. In the DS scheme, the real emission correction $d \sigma_R$ is subtracted
by the dipole term $d \sigma_{DP}$ before integration over the $(m+1)$-body phase space \footnote{$m=3$ for $ZZW$ production.}.
The dipole term approximates the divergent behavior of the real emission in all soft/collinear regions, which means
$(d \sigma_R - d \sigma_{DP})$ is finite and can be integrated in four dimensions directly. Then the NLO QCD corrected
cross section can be expressed as
\begin{eqnarray}
\sigma_{NLO}=
\sigma_{LO} +
\int\limits_{m}
\Bigl[
d \sigma_V + d \sigma_{PDF} + \int\limits_{1} d \sigma_{DP}
\Bigr] +
\int\limits_{m+1}
\Bigl[
d\sigma_R - d\sigma_{DP}
\Bigr],
\end{eqnarray}
where $d \sigma_V$ and $d \sigma_{PDF}$ are the virtual correction and the contribution of the PDF counterterms,
respectively. The integration over the real emission particle phase space $\int\limits_1 d\sigma^{DP}$ can be
computed analytically. As mentioned above, the IR divergences in $d \sigma_V$ can be canceled by those in
$d \sigma_{PDF}$ and $\int\limits_1 d\sigma^{DP}$. The integration of $(d \sigma_V + d \sigma_{PDF} + \int\limits_{1} d \sigma_{DP})$ over the $m$-body phase space can also be performed
numerically in four dimensions.

\par
The real emission corrections to the $pp \to ZZW^+ + X$ process are from the real gluon emission processes
$qq^{\prime} \to ZZW^+g$ $(qq^{\prime}=u\bar{d}, u\bar{s}, c\bar{d}, c\bar{s})$ and the real light-quark
emission processes $gq \to ZZW^+q^{\prime}$
$(qq^{\prime}= ud, \bar{d}\bar{u}, us, \bar{s}\bar{u}, cd, \bar{d}\bar{c}, cs, \bar{s}\bar{c})$.
In the case of real gluon emission two dipoles are needed as subtraction terms, while for the case of real light-quark
emission, only one subtraction term is needed. The analytical expressions for the dipoles are presented in Ref.\cite{ZZW-check2}.

\par
In the numerical calculation using the subtraction scheme, we will encounter the so-called missed binning problem \cite{Nagy},
because a huge positive weight from the real emission part and the corresponding huge negative weight from the
subtraction term may be filled into different histogram bins. In Ref.\cite{Nagy}, Zoltan Nagy introduces a parameter $\alpha$,
which decreases the size of dipole phase space, to distinct regions neighboring a singularity and regions without need of a
subtraction. We may suppress missed binning by taking proper value of $\alpha$.

\par
In Table \ref{tab1} we make the comparison between our numerical results for the $ZZW^+$ production and those from Refs.\cite{ZZW-check1}
and \cite{ZZW-check2} in the SM for some typical values of the factorization/renormalization scale. We take all the input parameter
values being the same as in Refs.\cite{ZZW-check1, ZZW-check2}. From Table \ref{tab1} we can see that our LO and NLO results are in agreement
with those provided in Ref.\cite{ZZW-check1}, and both the results of Ref.\cite{ZZW-check1} and ours are coincident at the $1\%$ level with those
given in Ref.\cite{ZZW-check2}.
\begin{table}[!hbp]
\small
\begin{center}
\begin{tabular}{|c|c|c|c|}
  \hline
    $\mu~[{\rm GeV}]$    &  $~{\rm data~source}~$          & ${\small ~~~\sigma_{LO}~[{\rm fb}]~~~}$  & ${\small ~~~\sigma_{NLO}~[{\rm fb}]~~~}$  \\
  \hline                 & {\small Ours }          & {\small 20.46(1)}                 & {\small 42.91(2)}     \\
{$\frac{3}{2}~m_Z$}      & {\small Ref.\cite{ZZW-check1}}   & {\small 20.42(3)}                & {\small 43.02(8)}     \\
                         & {\small Ref.\cite{ZZW-check2}}  & {\small 20.2(1)}                 & {\small 43.0(2)}      \\
  \hline                 & {\small Ours }          & {\small 20.31(1)}                & {\small 39.98(2)}     \\
{$2~m_Z+m_W$}            & {\small Ref.\cite{ZZW-check1}}   & {\small 20.30(3)}                & {\small 39.87(9)}     \\
                         & {\small Ref.\cite{ZZW-check2}}  & {\small 20.2(1)}                  & {\small 40.4(2)}      \\
  \hline                 & {\small Ours }          & {\small 20.30(1)}                 & {\small 39.83(2)}     \\
{$3~m_Z$}                & {\small Ref.\cite{ZZW-check1}}   & {\small 20.24(3)}                & {\small 39.86(7)}     \\
                         & {\small Ref.\cite{ZZW-check2}}  & {\small 20.0(1)}                 & {\small 39.7(2)}      \\
  \hline                 & {\small Ours }          & {\small 20.07(1)}                & {\small 37.40(1)}     \\
{$6~m_Z$}                & {\small Ref.\cite{ZZW-check1}}   & {\small 20.03(3)}                & {\small 37.39(7)}     \\
                         & {\small Ref.\cite{ZZW-check2}}  & {\small 19.7(1)}                  & {\small 37.8(2)}      \\
  \hline
\end{tabular}
\end{center}
\begin{center}
\begin{minipage}{15cm}
\caption{\label{tab1} The comparison between our SM results and
those from Refs.\cite{ZZW-check1} and \cite{ZZW-check2} for $ZZW^+$ production at the
$\sqrt{s}=14~{\rm TeV}$ LHC.}
\end{minipage}
\end{center}
\end{table}

\par
\section{Numerical results and discussions}
\par
\subsection{Input parameters }
\par
In our numerical calculations, the SM input parameters are taken as $\alpha_{{\rm ew}}^{-1}=137.035999074$,
$m_W=80.385~{\rm GeV}$, $m_Z=91.1876~{\rm GeV}$ \cite{PDG} and $m_H=126~{\rm GeV}$ \cite{Higgs-mass-1, Higgs-mass-2}. Since the LED
model is a low-enerty effective theory, it breaks down when $\sqrt{p_{G}^2} \gtrsim M_S$, where $p_{G}$ is the
four-momentum flowing through the KK graviton. For $ZZW$ production, $p_{G}^2$ is the invariant mass squared
of final $Z$-boson pair. The factorization and renormalization scales are set to be equal and the central value
of the factorization/renormalization scale is defined as $\mu_0 = 2 m_Z + m_W$.
In order to make reliable and viable phenomenological predictions, we take the hard
and conservative truncation scheme by applying the cut $M_{ZZ} < M_S$. For the initial state convolution, we
adopt CTEQ6L1 PDFs with $\Lambda_5^{LO}=165~{\rm MeV}$ and CTEQ6M PDFs \cite{CTEQ} with
$\Lambda_5^{\overline{MS}}=226~{\rm MeV}$ in the LO and NLO calculations, respectively.
The Cabibbo-Kobayashi-Maskawa (CKM) matrix elements are taken as
\begin{eqnarray}\label{CKM}
 V_{CKM} &=& \left(
\begin{array}{ccc}
    V_{ud} \ &  V_{us} \ &  V_{ub} \\
    V_{cd} \ &  V_{cs} \ &  V_{cb} \\
    V_{td} \ &  V_{ts} \ &  V_{tb} \\
\end{array}
    \right)=\left(
\begin{array}{ccc}
     ~~0.97425 \ &  0.22547 \ &  0 \\
    -0.22547 \ &  0.97425 \ &  0 \\
       0 \ &  0 \ &  1 \\
\end{array}  \right).
\end{eqnarray}

\par
Up to now, the ATLAS and CMS Collaborations at the LHC have not yet observed the signature of extra
spatial dimensions. All the experimental data are in good agreement with the SM predictions and
thus provide more severe constraints on the LED parameters. The ATLAS Collaboration provided lower
limits on $M_S$ at $95\%$ confidence level (CL) between $2.4~{\rm TeV}$ and $3.9~{\rm TeV}$ in
dilepton events at $\sqrt{s}=7~{\rm TeV}$ LHC depending on the choice of model, channel and prior.
After combining the dilepton and diphoton searches, the limits are in the range of $2.6-4.2~{\rm TeV}$ \cite{bound-atlas-1, bound-atlas-2}.
While the CMS Collaboration set the $M_S$ lower limits of up to $4.77~{\rm TeV}$ at $95\%$ CL in
dielectron events, depending on the number of extra dimensions and the validity range of the theory \cite{bound-cms}.
In discussing the LED effects, we apply two event selection schemes. In scheme (I) we collect all the
events without any cut on the final products, while in scheme (II) we only accept the events satisfying
the following event selection criteria:
\begin{eqnarray}
M_{ZZ} > 500~{\rm GeV},~~~~ p_T^{Z} > 100~{\rm GeV},~~~~ p_T^{W}>20~{\rm GeV}.
\end{eqnarray}
In our calculations the LED parameters are taken as $M_S=4.8~{\rm TeV}$, $d=3$ and adopt scheme (II) for event selection unless otherwise stated.

\subsection{Production cross section}
\par
In Table \ref{tab2} we present the integrated cross sections and the corresponding QCD $K$-factor in the SM
and the LED model, and the LED relative discrepancy $\delta$ which is defined as $\delta=(\sigma_{LED}-\sigma_{SM})/\sigma_{SM}$,
for the $ZZW^+$ and $ZZW^-$ productions in the the event selection scheme (I) at the $\sqrt{s}=14~{\rm TeV}$ LHC.
The LO relative deviations between the total cross sections predicted in the LED model and the SM for the $ZZW^+$ and $ZZW^-$
productions are $2.40\%$ and $1.23\%$, respectively. Like most of the TGB production processes, the LO cross
sections for the $ZZW^+$ and $ZZW^-$ productions are enhanced significantly by the NLO QCD corrections,
but the QCD corrections strongly reduce the LO LED relative discrepancies and make the NLO relative deviations
down to $1.03\%$ and $0.65\%$, respectively. In other words, the LO result overestimates the LED effect and
the NLO LED signal in scheme (I) is almost submerged in the SM background with our chosen parameters.
\begin{table}[!hbp]
\small
\begin{center}
  \begin{tabular}{|c|ccc|ccc|}
  \hline
  && $ZZW^+$ && & $ZZW^-$ & \\
  & ~~~~$\sigma_{LO}~(fb)$ & ~$\sigma_{NLO}~(fb)$  & ~~~~{\small $K$}~~~~ & ~~~~$\sigma_{LO}~(fb)$  & ~$\sigma_{NLO}~(fb)$  & ~~~~{\small $K$}~~~~ \\
  \hline
  SM &  {\small 18.29(1)} & {\small 36.84(2)} & {\small 2.014} & {\small 9.428(4)} & {\small 20.01(1)} & {\small 2.122} \\
  LED & {\small 18.73(1)} & {\small 37.22(2)} & {\small 1.987} & {\small 9.544(4)} & {\small 20.14(1)} & {\small 2.110} \\
  $\delta$ & {\small $2.40\%$} & {\small $1.03\%$} & $-$ & {\small $1.23\%$} & {\small $0.65\%$} & $-$ \\
  \hline
  \end{tabular}
\end{center}
\begin{center}
\begin{minipage}{15cm}
\caption{\label{tab2} Integrated cross sections in scheme (I) for the $pp \to ZZW^{\pm} + X$ processes in the
SM and the LED model at the $\sqrt{s}=14~{\rm TeV}$ LHC with $\mu=\mu_0=2m_Z+m_W$, $M_S=4.8~{\rm TeV}$ and $d=3$. }
\end{minipage}
\end{center}
\end{table}

\par
\par
In Table \ref{tab3} we provide the results by taking the event selection scheme (II) with the same input parameters as in
Table \ref{tab2}. The results in scheme (II) show that both the LO and NLO LED relative discrepancies between the SM and LED predictions increase significantly. The LO (NLO) LED relative deviations reach the values of $19.88\%$ ($8.05\%$) and $12.30\%$ ($5.63\%$) for the $ZZW^+$ and $ZZW^-$ productions, respectively. Here we see also that the LO prediction overestimates the LED relative deviation for the $ZZW$ production at the LHC.
\begin{table}[!hbp]
\small
\begin{center}
  \begin{tabular}{|c|ccc|ccc|}
  \hline
  && $ZZW^+$ && & $ZZW^-$ & \\
  & ~~~~$\sigma_{LO}~(fb)$ & ~$\sigma_{NLO}~(fb)$ & ~~~~{\small $K$}~~~~ & ~~~~$\sigma_{LO}~(fb)$  & ~$\sigma_{NLO}~(fb)$  & ~~~~{\small $K$}~~~~ \\
  \hline
  SM &  {\small 2.062(1)}       & {\small 4.372(6)}  & {\small 2.120} & {\small 0.878(1)}       & {\small 2.025(4)}  & {\small 2.306} \\
  LED & {\small 2.472(2)}       & {\small 4.724(7)}  & {\small 1.911} & {\small 0.986(1)}      & {\small 2.139(4)}  & {\small 2.169} \\
  $\delta$ & {\small $19.88\%$}      & {\small $8.05\%$} & $-$ & {\small $12.30\%$}      & {\small $5.63\%$} & $-$ \\
  \hline
  \end{tabular}
\end{center}
\begin{center}
\begin{minipage}{15cm}
\caption{\label{tab3} Integrated cross sections in scheme (II) for the $pp \to ZZW^{\pm} + X$ processes in the
SM and the LED model at the $\sqrt{s}=14~{\rm TeV}$ LHC with $\mu=\mu_0=2m_Z+m_W$, $M_S=4.8~{\rm TeV}$ and $d=3$. }
\end{minipage}
\end{center}
\end{table}

\subsection{Scale dependence}
\par
\begin{figure}
\begin{center}
\includegraphics[scale=0.45]{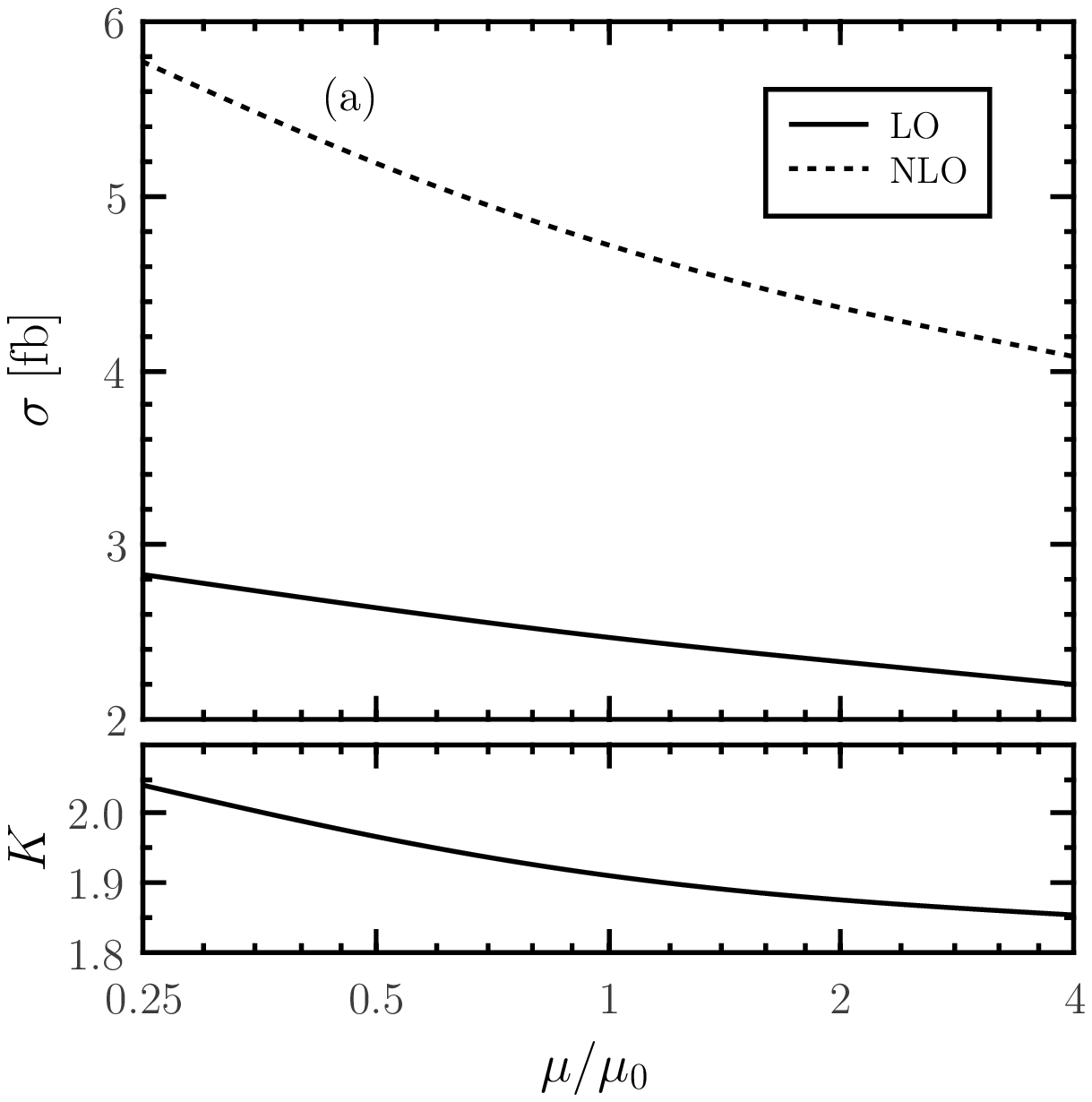}%
\includegraphics[scale=0.45]{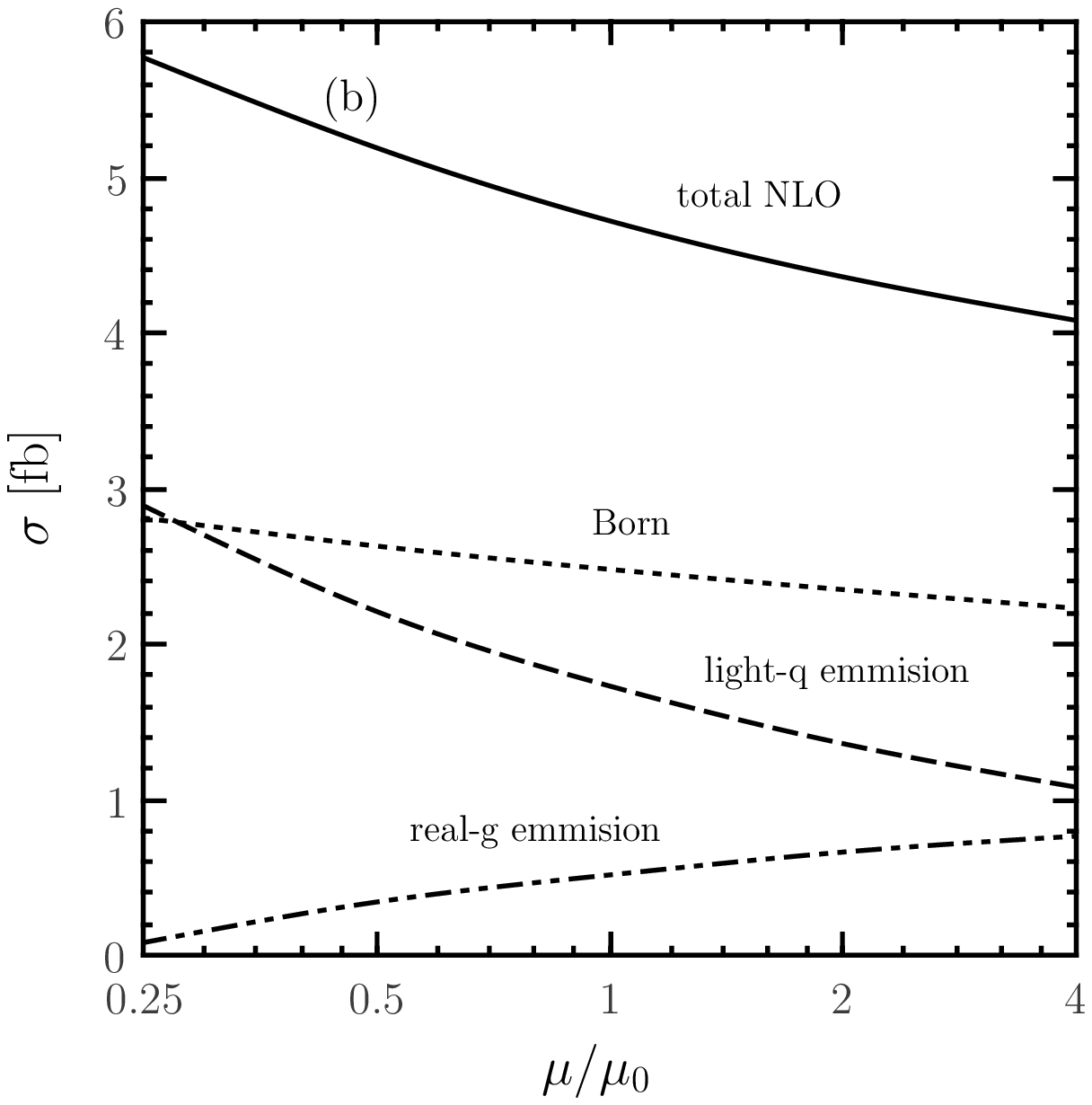}%
\caption{\label{scale-Wplus} Scale dependence of the LO and NLO QCD corrected cross sections for the $pp \to ZZW^+ + X$ process
at the $\sqrt{s}=14~{\rm TeV}$ LHC in the LED model. (a) Integrated LO, NLO QCD corrected cross sections and the corresponding
$K$-factors. (b) Different contribution parts to the NLO total cross section. }
\end{center}
\end{figure}
\begin{figure}
\begin{center}
\includegraphics[scale=0.45]{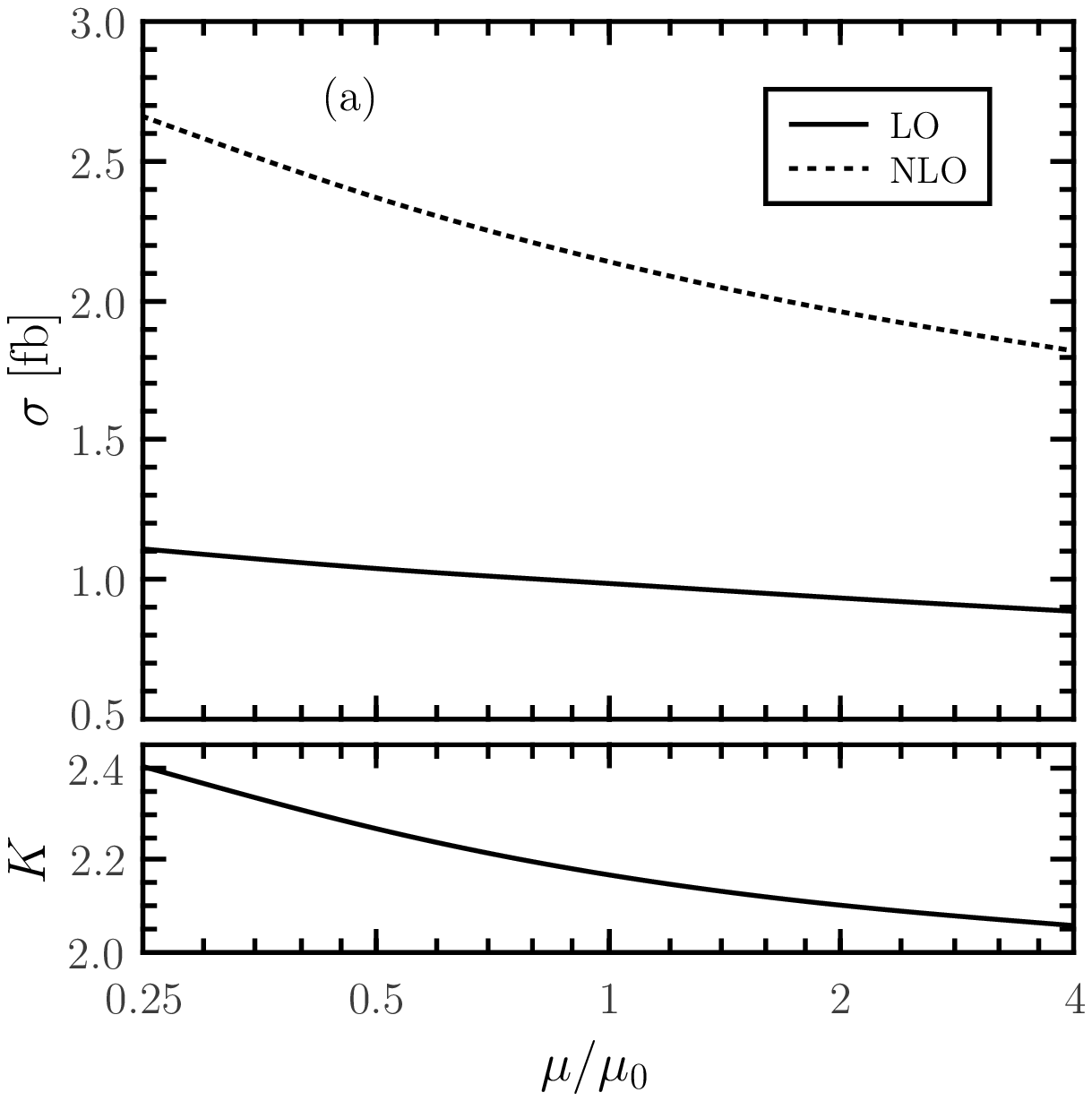}%
\includegraphics[scale=0.45]{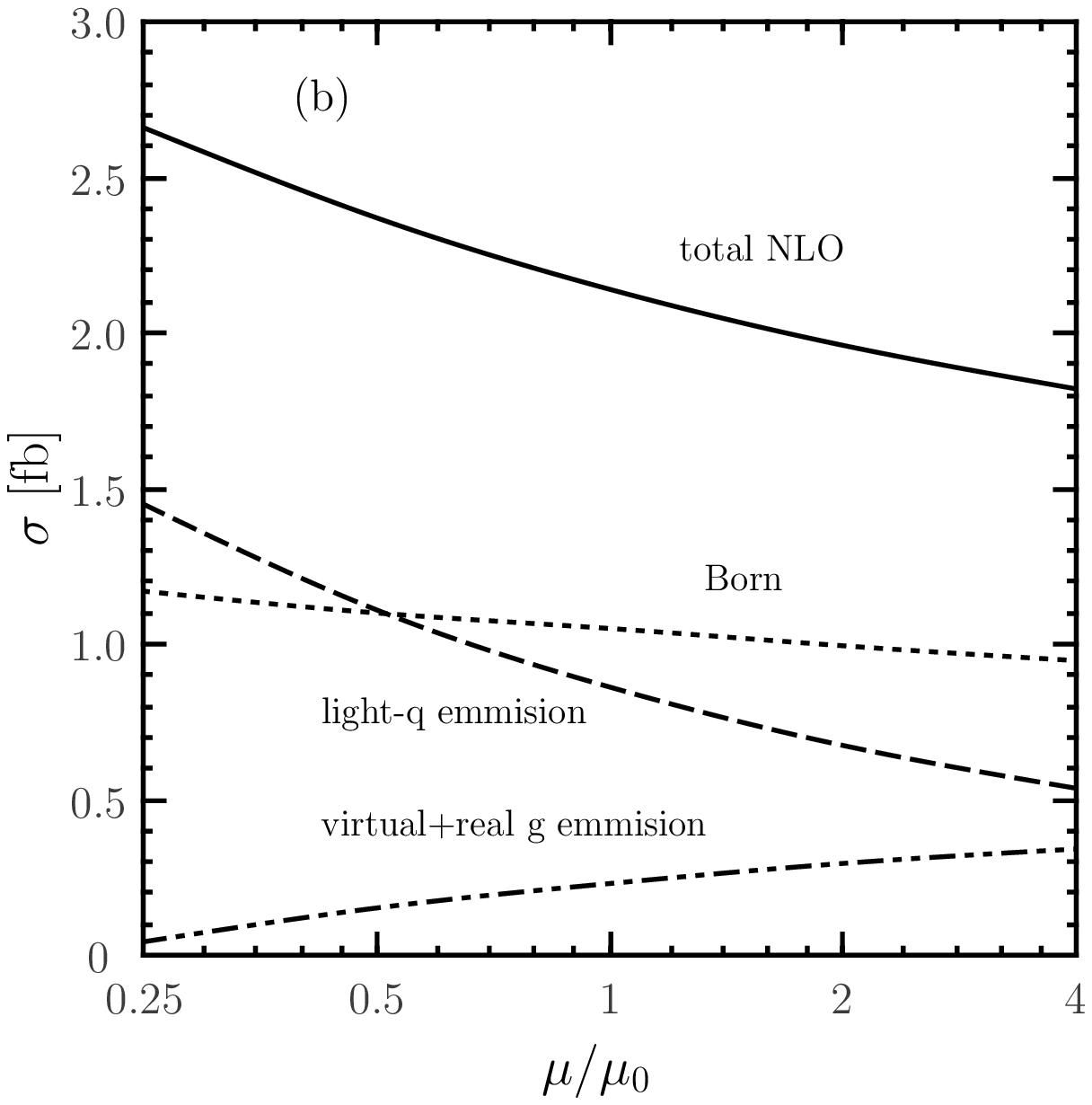}%
\caption{\label{scale-Wminus} Scale dependence of the LO and NLO QCD corrected cross sections for the $pp \to ZZW^- + X$ process
at the $\sqrt{s}=14~{\rm TeV}$ LHC in the LED model. (a) Integrated LO, NLO QCD corrected cross sections and the corresponding
$K$-factors. (b) Different contribution parts to the NLO total cross section. }
\end{center}
\end{figure}

We depict the LO, NLO QCD corrected total cross sections and the corresponding $K$-factors for the $ZZW^+$ and $ZZW^-$
productions in the LED model at the $\sqrt{s}=14~{\rm TeV}$ LHC as functions of the factorization/renormalization scale
in Fig.\ref{scale-Wplus}(a) and Fig.\ref{scale-Wminus}(a), respectively. There we set $\mu \equiv \mu_f=\mu_r$ and take the LED parameters
as $M_S=4.8~{\rm TeV}$ and $d=3$. We can read out from Fig.\ref{scale-Wplus}(a) and Fig.\ref{scale-Wminus}(a) that the $K$-factors range
from $2.04$ to $1.85$ and from $2.40$ to $2.06$ for the $ZZW^+$ and $ZZW^-$ productions, respectively, with the increment
of $\mu$ from $0.25\mu_0$ to $4\mu_0$. We also find that the LO cross section underestimates the scale uncertainty,
because the $ZZW^+$ and $ZZW^-$ productions at the LO are pure electroweak processes, and the LO scale uncertainty is apparently
only related to the PDFs. While the NLO scale uncertainty is related to both the factorization scale and the renormalization
scale, and is enhanced obviously due to $\alpha_s(\mu_r)$ appearing at the NLO.

\par
For demonstrating the main origin of the scale uncertainty, we plot the NLO QCD corrected total cross sections, Born
contributions, real light-quark emission and virtual plus real gluon emission correction components for the $ZZW^+$ and $ZZW^-$
productions in the LED model at the $\sqrt{s}=14~{\rm TeV}$ LHC versus the scale $\mu$ in Fig.\ref{scale-Wplus}(b) and Fig.\ref{scale-Wminus}(b),
respectively. From these figures we can see that the NLO scale uncertainties for both the $ZZW^+$ and the $ZZW^-$ production
processes are mainly induced by the real light-quark emission corrections, which originate from the gluon-initiated subprocesses,
and are mainly responsible for the large value of $K$-factor.
\par

\subsection{LED parameter dependence }
\par
In Table \ref{tab4}, we list the LO, NLO QCD corrected integrated cross sections and the corresponding $K$-factors
for the $ZZW^+$ and $ZZW^-$ productions in the LED model at the $14~{\rm TeV}$ LHC for some typical values of $M_S$
and $d$. We can see that the integrated cross section in the LED model decreases with the increment of $M_S$ and
approaches to the corresponding SM prediction. On the other hand, for a fixed value of $M_S$ the deviation between
the predictions in the LED model and the SM increases when the value of $d$ becomes smaller. It can be ascribed to
the fact that the contribution of the interchanging KK-graviton is reduced with the increment of $M_S$ and/or $d$ (see Eq.(\ref{Res})), as
shown explicitly in the KK-graviton propagator listed in Table \ref{tab0}.
\begin{table}
\small
\begin{center}
\begin{tabular}{|c|ccc|ccc|ccc|}
  \hline
   ${\small M_S}$                      & \multicolumn{3}{|c|}{\small $d~=~3$}
                       & \multicolumn{3}{|c|}{\small $d~=~4$}
                       & \multicolumn{3}{|c|}{\small $d~=~5$} \\
    ${\small ({\rm TeV})}$   &
                {$\sigma_{LO}~(fb)$}    &{$\sigma_{NLO}~(fb)$}  & {\small $K$}  &     
                {$\sigma_{LO}~(fb)$}    &{$\sigma_{NLO}~(fb)$}  & {\small $K$}  &     
                {$\sigma_{LO}~(fb)$}    &{$\sigma_{NLO}~(fb)$}  & {\small $K$}     \\ 
  \hline
      $5$                 & {\small 2.374(2)}       & {\small 4.649(5)}  & {\small 1.959}
                          & {\small 2.276(1)}       & {\small 4.556(6)}  & {\small 2.002}
                          & {\small 2.220(1)}       & {\small 4.508(6)}  & {\small 2.030}     \\
     $5.5$                & {\small 2.228(1)}       & {\small 4.527(6)}  & {\small 2.032}
                          & {\small 2.172(1)}       & {\small 4.468(6)}  & {\small 2.057}
                          & {\small 2.142(1)}       & {\small 4.439(6)}  & {\small 2.072}     \\
     $6$                  & {\small 2.155(1)}       & {\small 4.461(7)}  & {\small 2.070}
                          & {\small 2.122(1)}       & {\small 4.436(5)}  & {\small 2.091}
                          & {\small 2.104(1)}       & {\small 4.411(6)}  & {\small 2.097}     \\
     $6.5$                & {\small 2.117(1)}       & {\small 4.435(6)}  & {\small 2.095}
                          & {\small 2.097(1)}       & {\small 4.418(5)}  & {\small 2.107}
                          & {\small 2.085(1)}       & {\small 4.404(6)}  & {\small 2.112}     \\
  \hline
\end{tabular}
\end{center}
\small
\begin{center}
\begin{tabular}{|c|ccc|ccc|ccc|}
  \hline
   ${\small M_S}$                      & \multicolumn{3}{|c|}{\small $d~=~3$}
                       & \multicolumn{3}{|c|}{\small $d~=~4$}
                       & \multicolumn{3}{|c|}{\small $d~=~5$} \\
    ${\small ({\rm TeV})}$   &
                {$\sigma_{LO}~(fb)$}    &{$\sigma_{NLO}~(fb)$}  & {\small $K$}  &
                {$\sigma_{LO}~(fb)$}    &{$\sigma_{NLO}~(fb)$}  & {\small $K$}  &
                {$\sigma_{LO}~(fb)$}    &{$\sigma_{NLO}~(fb)$}  & {\small $K$}     \\
  \hline
      $5$                 & {\small 0.961(1)}       & {\small 2.115(3)}  & {\small 2.201}
                          & {\small 0.934(1)}       & {\small 2.085(3)}  & {\small 2.232}
                          & {\small 0.919(1)}       & {\small 2.065(3)}  & {\small 2.247}     \\
     $5.5$                & {\small 0.923(1)}       & {\small 2.075(3)}  & {\small 2.247}
                          & {\small 0.908(1)}       & {\small 2.059(3)}  & {\small 2.269}
                          & {\small 0.899(1)}       & {\small 2.047(3)}  & {\small 2.276}     \\
     $6$                  & {\small 0.904(1)}       & {\small 2.056(3)}  & {\small 2.273}
                          & {\small 0.895(1)}       & {\small 2.046(3)}  & {\small 2.288}
                          & {\small 0.889(1)}       & {\small 2.040(3)}  & {\small 2.294}     \\
     $6.5$                & {\small 0.894(1)}       & {\small 2.050(3)}  & {\small 2.293}
                          & {\small 0.888(1)}       & {\small 2.038(3)}  & {\small 2.295}
                          & {\small 0.885(1)}       & {\small 2.036(3)}  & {\small 2.301}     \\
  \hline
\end{tabular}
\end{center}
\begin{center}
\begin{minipage}{15cm}
\caption{\label{tab4} The LO, NLO QCD corrected cross sections and the corresponding $K$-factors for the
$pp \to ZZW^+ + X$ (upper table) and $pp \to ZZW^- + X$ (lower table) processes in the LED model
with $\mu=\mu_0$ at the $\sqrt{s}=14~{\rm TeV}$ LHC for some typical values of $M_S$ and $d$. }
\end{minipage}
\end{center}
\end{table}

\subsection{LED effects on differential distributions}
\par
To describe the LED effect on the differential distribution with respect to a kinematic variable $x$,
we introduce a quantity named LED relative discrepancy defined as
$\delta(x) = \left(\frac{d\sigma_{LED}}{dx}-\frac{d\sigma_{SM}}{dx}\right) \Big/\frac{d\sigma_{SM}}{dx}$.
Due to the $CP$-conservation, the difference between the observables for the $ZZW^+$ production and
those for the $ZZW^-$ production at the LHC only comes from the different PDFs of the incoming partons.
Therefore, we provide only the kinematic distributions for the $pp \to ZZW^+ + X$ process as
a representative in further discussion.

\par
\begin{figure}[htbp]
\begin{center}
\includegraphics[scale=0.45]{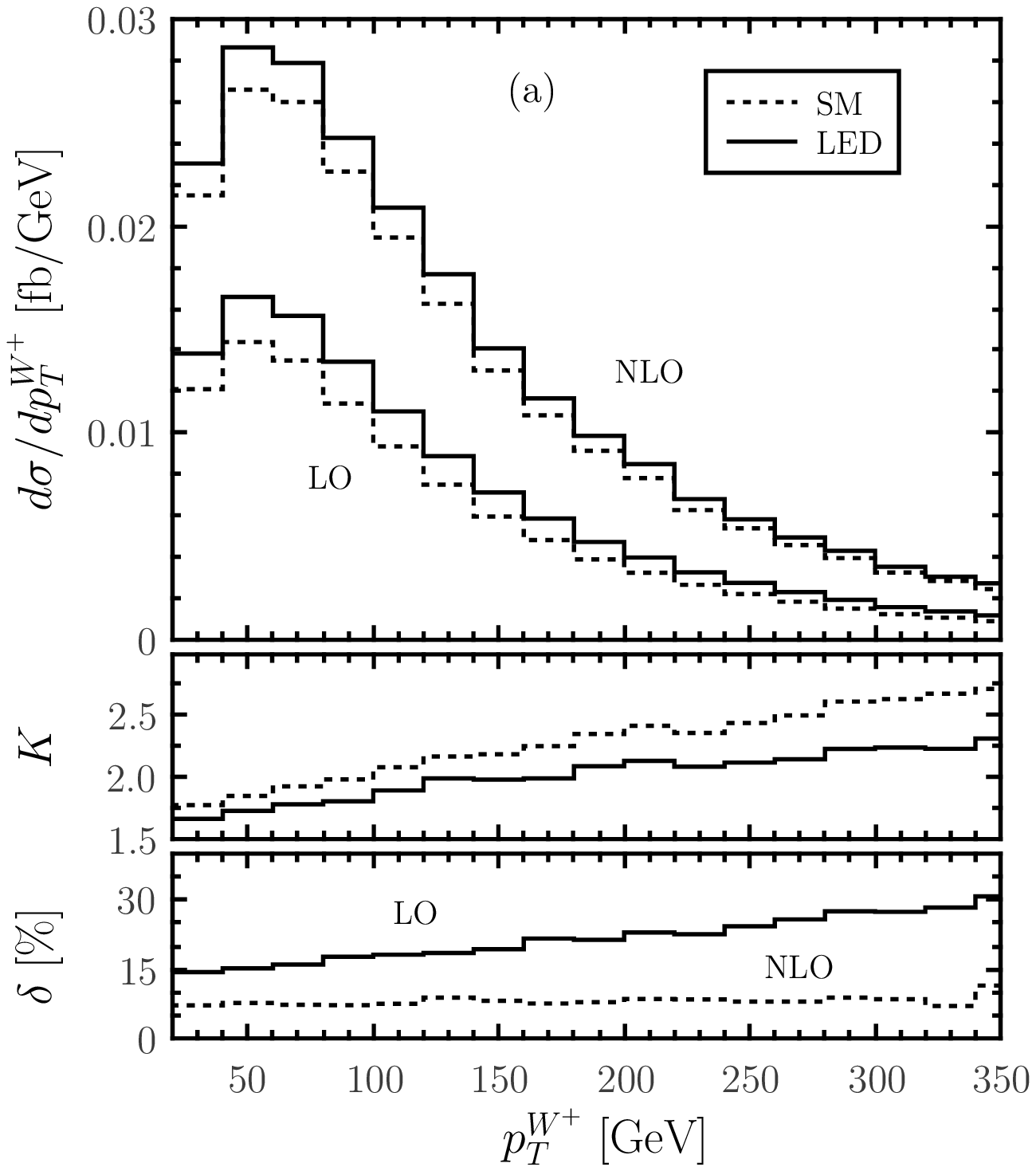}%
\includegraphics[scale=0.45]{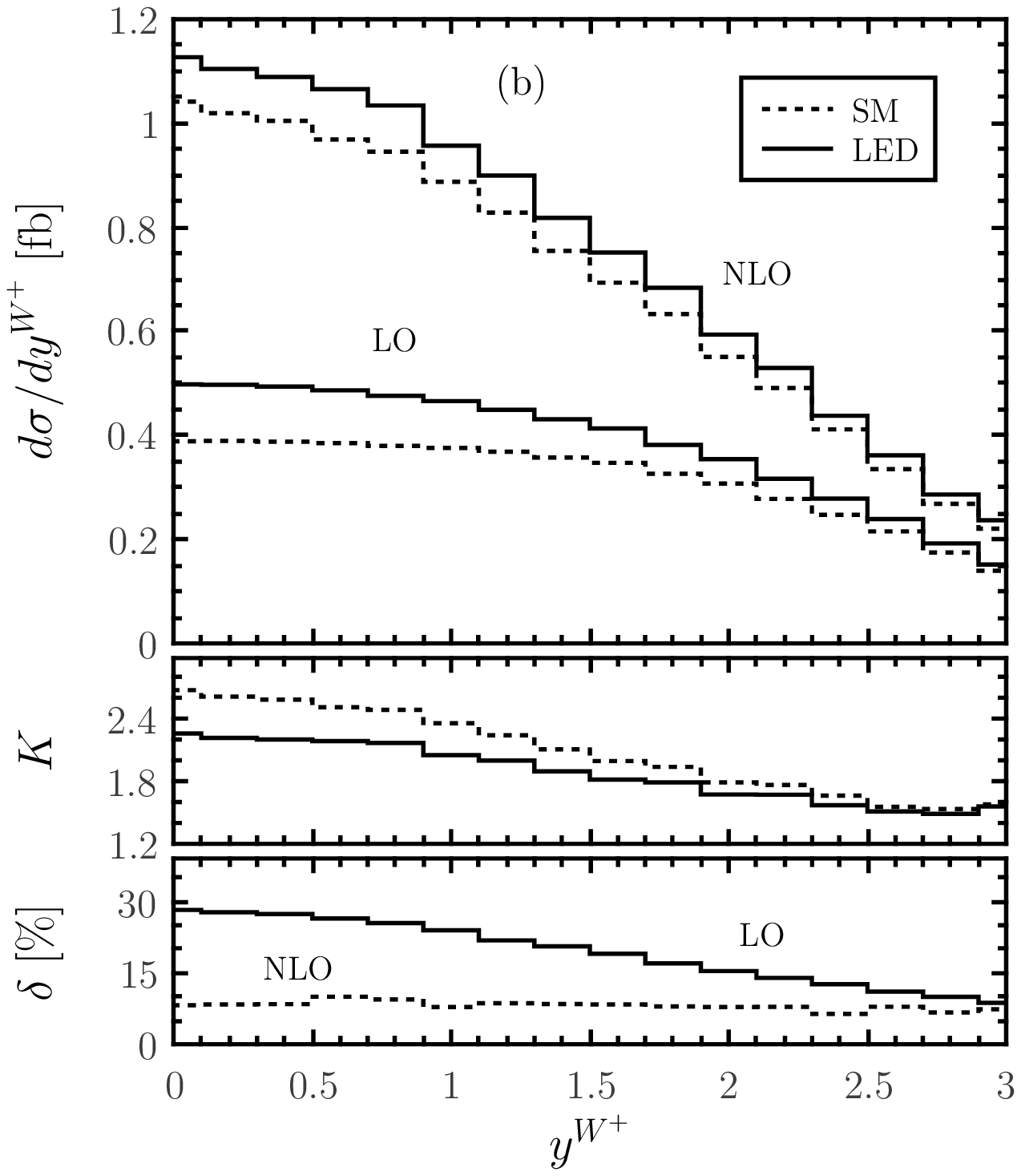}%
\caption{ \label{diff-W}
The LO and NLO QCD corrected kinematic distributions of final $W^+$-boson for the $ZZW^+$ production at the $\sqrt{s}=14~{\rm TeV}$ LHC in both the SM and the LED model, and the corresponding $K$-factors and the LED relative discrepancies. (a) $p_T^{W^+}$ distributions. (b) $y^{W^+}$ distributions. }
\end{center}
\end{figure}

Due to the symmetric feature of the rapidity $y$ we study the behaviour of the rapidity distribution only in the positive rapidity region ($y \in [0,~3]$) in following discussions. In Figs.\ref{diff-W}(a, b), we provide the LO, NLO QCD corrected transverse momentum and rapidity distributions of final $W^+$-boson for the $ZZW^+$ production at the $\sqrt{s}=14~{\rm TeV}$ LHC in both the SM and the LED model. The corresponding  $K$-factors and the LED relative discrepancies are also plotted in these figures. As shown in Fig.\ref{diff-W}(a), the LED effect becomes larger with the increment of $p_T^{W^+}$. The LO LED relative discrepancy $\delta_{LO}(p_T^{W^+})$ ranges from $14.3\%$ to $27.5\%$ in the region of $20~{\rm GeV} < p_T^{W^+} < 300~{\rm GeV}$, while $\delta_{NLO}(p_T^{W^+})$, which is heavily suppressed by the NLO QCD corrections, goes up from $7.15\%$ to $8.82\%$ in the same $p_T^{W^+}$ region. From Fig.\ref{diff-W}(b) we can see that $\delta_{LO}(y^{W^+})$ in the range of $y^{W^+}\in [0,~3]$ is larger than $8\%$ and has its maximum of about $28.2\%$ at $y^{W^+} = 0$, while $\delta_{NLO}(y^{W^+})$, which is remarkably suppressed, ranges from $6.38\%$ to $9.97\%$ in the region of $0 < y^{W^+} < 3$. Both the two figures show that the NLO QCD correction is significant, and the $K$-factors exceed $1.5$ in both the SM and the LED model in the plotted $p_T^{W^+}$ and $y^{W^+}$ regions.
\par

\par
\begin{figure}[htbp]
\begin{center}
\includegraphics[scale=0.45]{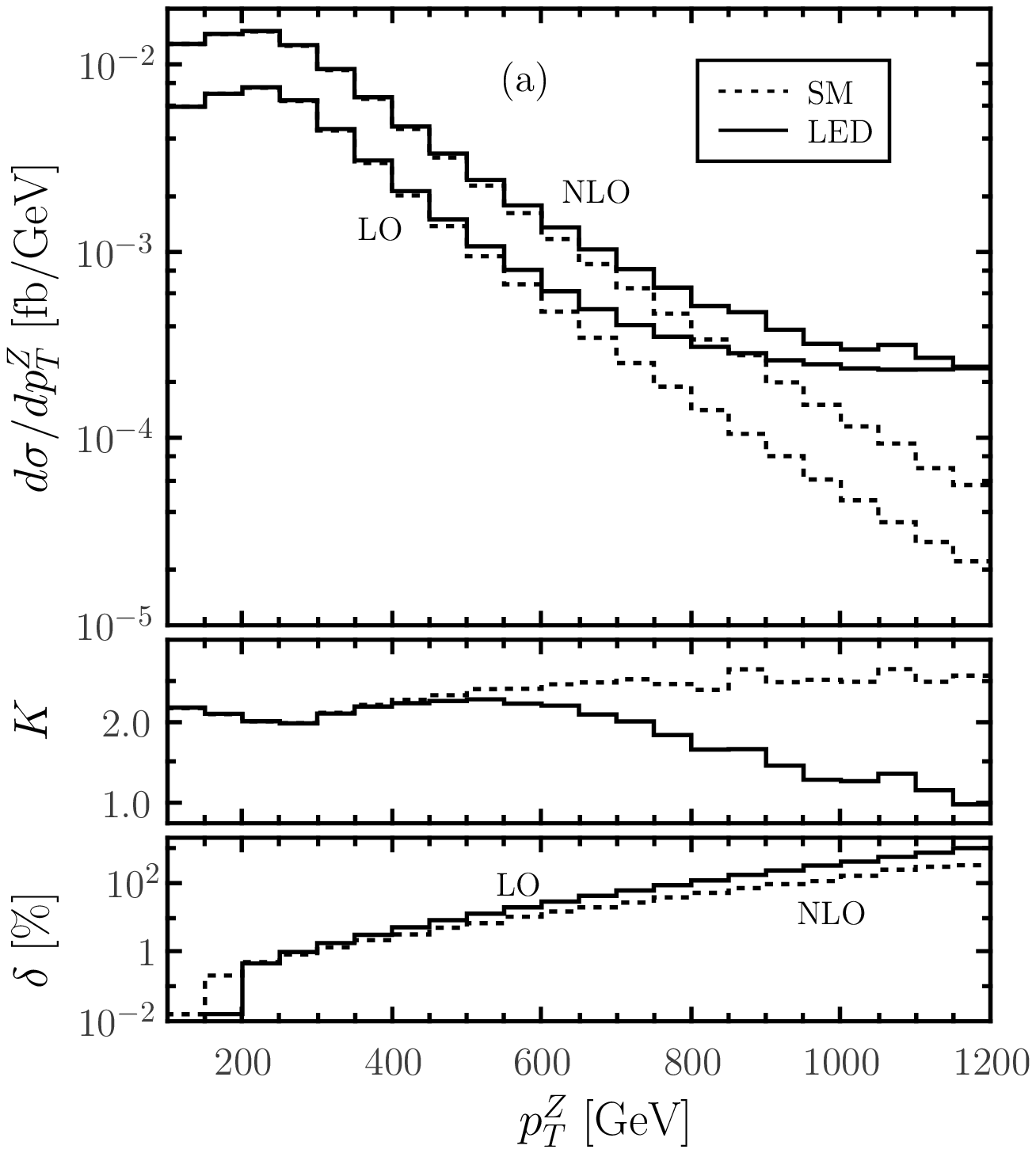}%
\includegraphics[scale=0.45]{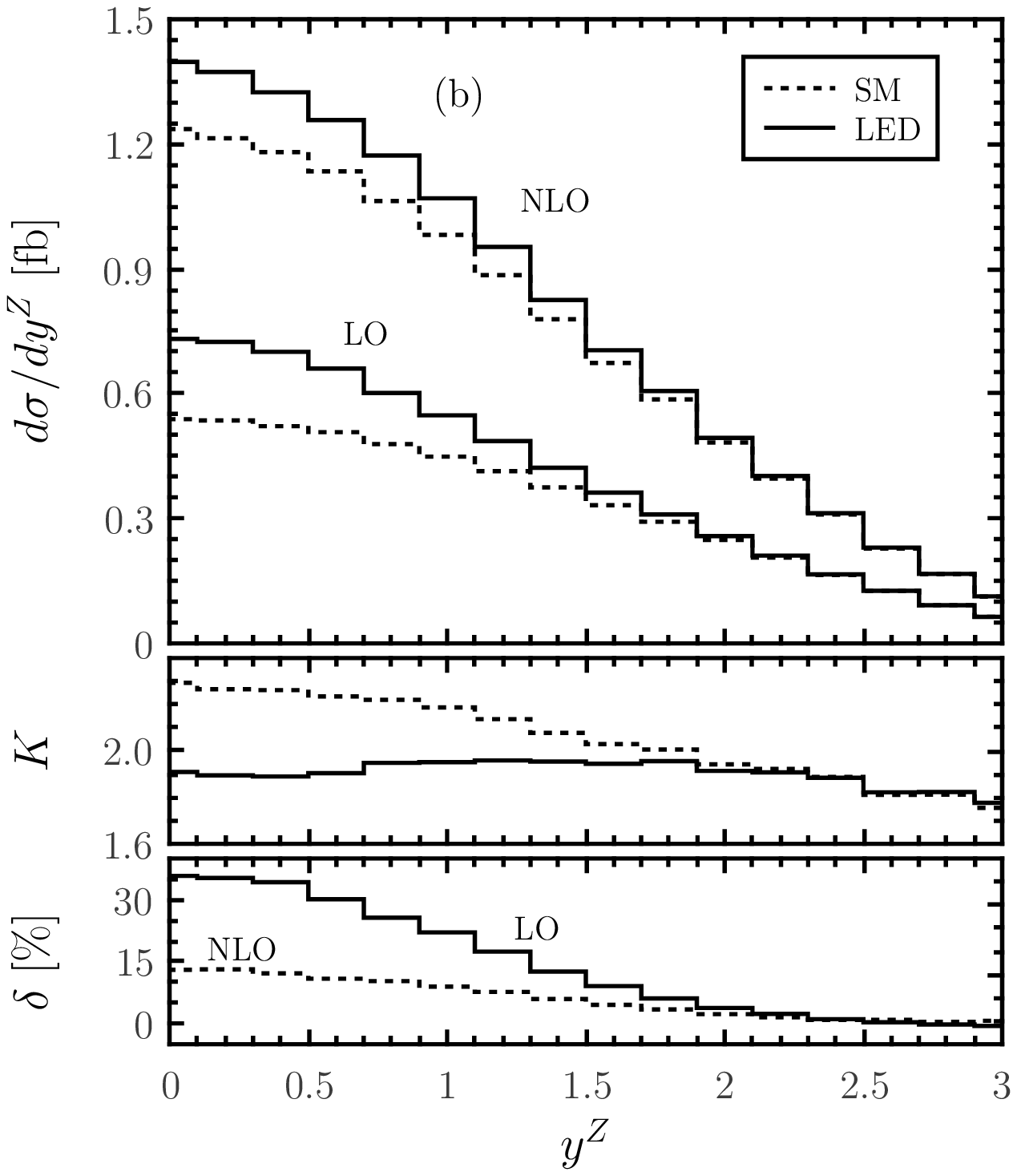}%
\caption{ \label{diff-Z}
The LO and NLO QCD corrected kinematic distributions of final $Z$-bosons for the $ZZW^+$ production at the $\sqrt{s}=14~{\rm TeV}$ LHC in both the SM and the LED model, and the corresponding $K$-factors and the LED relative discrepancies. (a) $p_T^{Z}$ distributions. (b) $y^{Z}$ distributions.}
\end{center}
\end{figure}

The transverse momentum and rapidity distributions of final two $Z$-bosons for the $pp \to ZZW^+ + X$ process at the $\sqrt{s}=14~{\rm TeV}$ LHC, are depicted in Fig.\ref{diff-Z}(a) and Fig.\ref{diff-Z}(b), respectively. In these figures we pick $p_T^{Z}$ and $y^{Z}$ of each of the two identical $Z$-bosons as an entry in the histograms, then the final differential cross section should be multiplied by  $\frac{1}{2}$. The corresponding $K$-factors and the LED relative discrepancies are depicted in the lower plots of Figs.\ref{diff-Z}(a, b). We can read out from Fig.\ref{diff-Z}(a) that both the LO and the NLO LED relative discrepancies are less that $10\%$ in the region of $p_T^Z \in [100,~400]~{\rm GeV}$, but they grow up quickly and become to be very large, when $p_T^{Z}$ goes up beyond $400~{\rm GeV}$. That is because the LED contributions induced by the KK gravitons enhance the differential distributions. As shown in Fig.\ref{diff-Z}(b), the $y^{Z}$ distributions are similar to the $y^{W^+}$ distributions shown in Fig.\ref{diff-W}(b), and the significant LED contributions are concentrated in low $y^{Z}$ region. The LO and NLO QCD corrected LED relative discrepancies reach their maxima of about $35.77\%$ and $13.05\%$ at $y^{Z}=0$, respectively.

\par
\begin{figure}[!htbp]
\begin{center}
\includegraphics[scale=0.45]{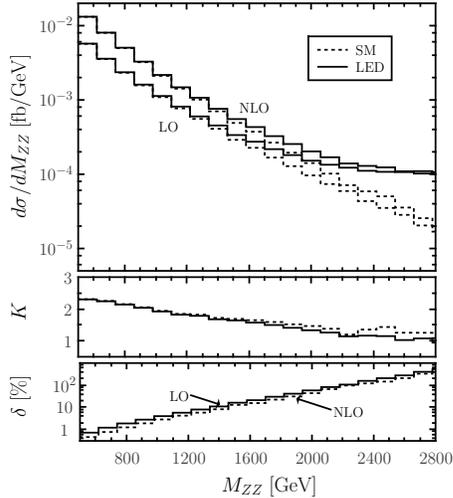}%
\caption{ \label{diff-invZmass}
The LO and NLO QCD corrected distributions of the $Z$-boson pair invariant mass for the $ZZW^+$ production at the $\sqrt{s}=14~{\rm TeV}$ LHC in both the SM and the LED model, and the corresponding $K$-factors and the LED relative discrepancies.   }
\end{center}
\end{figure}

In Fig.\ref{diff-invZmass} we present the LO and NLO QCD corrected distributions of the $Z$-boson pair invariant mass $M_{ZZ}$ for the $pp \to ZZW^+ + X$ process at the $\sqrt{s}=14~{\rm TeV}$ LHC in both the SM and the LED model, and the corresponding $K$-factors and the LED relative discrepancies are shown in the lower plots. It shows that the LED effect increases rapidly as the increment of $M_{ZZ}$, while is very small in relatively low $M_{ZZ}$ region. This behavior of the $M_{ZZ}$ distribution can be interpreted as that the contribution of the KK-graviton propagator increases distinctly with the increment of $M_{ZZ}$ since the KK graviton interacts directly with the final $Z$-boson pair (see Eq.(\ref{Res}) and Fig.\ref{Feyndiag}). When $M_{ZZ}$ goes beyond $2.5~{\rm TeV}$, $K$-factors approach to 1 in both SM and the LED model.

\par
All the above kinematic distributions show that the LED effect could be significant for $ZZW$ production at the $14~{\rm TeV}$ LHC by adopting proper event selection criteria, particularly in
the high $p_T$, central rapidity $y$ and large $M_{ZZ}$ regions, the LED effect becomes to be evidently large. We see that after including the NLO QCD corrections, the LED effect is reduced remarkably. We conclude that the LO result for the $ZZW$ production at the $14~{\rm TeV}$ LHC overestimates the LED effect.

\par
\section{Summary}
\par
We investigate the LED effect induced by the KK gravitons on the $ZZW$ production at the $\sqrt s=14~{\rm TeV}$ LHC up to the QCD NLO. We also study the factorization/renormalization scale dependence of the total cross section and show that the LO prediction underestimates the scale uncertainty. Some kinematic distributions are provided in both the SM and the LED model. Our numerical results demonstrate that the NLO QCD corrections are sizeable and reduce the LED effect remarkably, and the NLO QCD correction and the LED effect are strongly related to phase space. We conclude that the LO result overestimates the LED effect and is insufficient to provide a believable theoretical prediction in the LED model, and the NLO LED relative discrepancy of the total cross section could become sizable for the $ZZW$ production by adopting proper event selection scheme.

\par
\noindent{\large\bf Acknowledgments:} Thanks for the help and support of Supercomputing center of USTC in our numerical calculations. This work was supported in part by the National Natural Science Foundation of China (Grants. No.11275190, No.11375008, No.11375171).

\end{document}